\documentclass[aip,cha,preprint,numerical]{revtex4-1}
\usepackage{amsmath}
\usepackage{amssymb}
\usepackage{bm}
\usepackage{graphicx}
\usepackage{psfrag}
\usepackage{dcolumn}
\usepackage{bm}
\usepackage{natbib}		
\usepackage{color}

\newcommand {\vp} {\varphi}

\newcommand {\tcb} {}

\setcitestyle{numbers,open={[},close={]}}

\begin{document}
\title{Describing dynamics of driven multistable 
oscillators with phase transfer curves}

\author{Evgeny Grines}
\affiliation{Lobachevsky State University of Nizhny Novgorod, 
Department of Control Theory and Dynamics of Systems, 23, Prospekt Gagarina, Nizhny Novgorod, 603950, Russia}
\author{Grigory Osipov}
\affiliation{Lobachevsky State University of Nizhny Novgorod, 
Department of Control Theory and Dynamics of Systems, 23, Prospekt Gagarina, Nizhny Novgorod, 603950, Russia}
\author{Arkady Pikovsky}
\affiliation{Institute of Physics and Astronomy, University of Potsdam, 
Karl-Liebknecht-Str. 24/25, 14476 Potsdam-Golm, Germany}
\affiliation{Lobachevsky State University of Nizhny Novgorod, 
Department of Control Theory and Dynamics of Systems, 23, 
Prospekt Gagarina, Nizhny Novgorod, 603950, Russia}

\date{\today}

\begin{abstract}
Phase response curve is an important tool in studies of stable
self-sustained oscillations; it describes a phase shift under action of an 
external perturbation. We consider multistable oscillators with several stable limit cycles.
Under a perturbation, transitions from one oscillating mode to another one may occur. We 
define phase transfer curves to describe the phase shifts at such transitions. This allows
for a construction of one-dimensional maps that characterize periodically
kicked multistable oscillators. We show, that these maps are good approximations
of the full dynamics for large periods of forcing.
\end{abstract}

\pacs{
  05.45.Xt %	Synchronization; coupled oscillators \\
  }
\keywords{}%Use showkeys class option if keyword
                              %display desired
\maketitle

\begin{quotation}
For many practical applications it is important to know how
oscillators respond to perturbations. If the oscillator is stable, 
after a perturbation it returns to its oscillating mode, only the phase
is shifted. This is quantified by a phase response curve, which is an important tool
to study dynamics of forced and coupled oscillators. Quite often there exist 
several stable oscillating modes, one speaks in this case on multistability.
For multistable oscillators, external action may result in a switching from
one mode to another one. We extend the concept of phase response curve on this case
by introducing a phase transfer curve, describing dependence of the phase of
the target mode on the phase of the source one. 
\end{quotation}

\section{Introduction}
The concept of phase response curves (PRC) is widely used in the theory
of oscillations to describe sensitivity of limit cycle oscillations
to external actions~\cite{Canavier-06,Pikovsky-Rosenblum-Kurths-01,Izhikevich-07}.
Moreover, because PRCs can be rather easy measured in experiments,
 they find broad applications in studies of oscillating processes in 
 life systems, where
 equations for the oscillators are hardly available, see e.g. 
 \cite{Abramovich-Sivan_Akselrod-98,Hilaire_etal-12,Ikeda-82,Ikeda-Yoshizawa-Sato-83,%
 Khalsa_etal-03,Kralemann_et_al-13,Lengyel_et_al-05,Galan-Ermentrout-Urban-05}. 
 The form
 of the PRC is important for controlling the oscillator by external 
 forcing~\cite{Harada_etal-10,Zlotnik_etal-13,Pikovsky-15b}. One of practical 
 applications is optimal re-adjustment of the phase of circadian rhythms of 
 humans~\cite{Eastman-Burgess-09}. \tcb{Here, the concept of PRC (sometimes one
 speaks on Phase Transition Curves~\cite{Johnson1992,OhFuTo2015,WeiLee2001})
 is widely used to describe the effect of different stimuli (such as light pulses, 
 temperature pulses, or pulses of drugs or
chemicals) on the circadian  rhythm.} Another application is suppression
 of Parkinson's tremor by phase resetting~\cite{Popovych-Tass-12}.
 The concept of PRC has been also  generalized on chaotic and stochastic
 oscillators~\cite{Schwabedal-Pikovsky-Kralemann-Rosenblum-12,Schwabedal-Pikovsky-13}.
 
 The goal of this paper is to generalize the concept of phase response curves 
 on multistable
 limit-cycle
 oscillators. There, of course, also a standard PRC can be defined, 
 if the perturbation does not
 result in a transition from one limit cycle to another one. In the case such 
 a transition occurs, we
 define a Phase Transfer Curve (PTC) which provides dependence of the new 
 phase (on the target limit cycle) on the old phase (on the source limit cycle). 
 \tcb{[This concept should not be mixed with the concept of Phase Transition Curves 
 in circadian rhythms,
 where the source and the target cycles coinside.]}
 Below, as a basic example, we will consider the simplest case of bistable oscillations. 
 Furthermore, we will
 demonstrate how well the dynamics of a periodically forced bistable system
 can be described solely in terms of PRCs and PTCs, in comparison with the full 
 description that involves
 also the amplitudes.

\section{Phases for multistable periodic motions} 
\label{sec:pb}

The concept of the phase for
a system with  periodic self-sustained oscillations is based on the notion of 
isochrons~\cite{Guckenheimer-75}.
Consider an autonomous continuous-time dynamical system with variables $\mathbf{x}$.
On a limit cycle $\mathbf{x}^{lc}$ with period
$T$ and frequency $\omega=2\pi/T$, the phase is defined as a $2\pi$-periodic 
variable rotating 
uniformly in time $\dot\varphi(\mathbf{x}^{lc})=\omega$. For an asymptotically 
stable limit cycle with
a basin of attraction $U$, one can extend the definition of the phase 
to the whole basin. Consider a stroboscopic
map $\mathbf{x}\to\mathbf{x}$ defined for the time interval $T$, i.e. 
exactly for 
the period of oscillations. For this map,
all the points on the limit cycle (parametrised by their phases $\varphi$)
are fixed points, and all the points in the basin $U$ converge to one 
of the points on the limit cycle. Isochrons $I(\varphi)$ are defined as the 
stable manifolds of the fixed points on the limit cycle. Clearly, these manifolds 
foliate the whole basin $U$,
thus providing the phase $\varphi=\Phi(\mathbf{x})$ for all points in $U$. 
An isochron $I(\varphi)$
is a set of all points, which converge, under the time evolution from some 
initial moment $t_0$, 
to a point $\mathbf{x}(t)$ on the limit cycle that has 
the phase $\varphi$ at time $t=t_0$.

Quite often one uses a definition of \textit{asymptotic phase} which is equivalent to
one above. Evolution of 
each point $\mathbf{x}(0)$ in the basin of attraction of the limit cycle brings 
it to the limit cycle,
i.e. there exists a point $\mathbf{x}^{lc}(0)$ such 
that $|\mathbf{x}(t)-\mathbf{x}^{lc}(t)|\to 0$ 
as $t\to\infty$. Then one defines $\Phi(\mathbf{x}(0))=\varphi(\mathbf{x}^{lc}(0))$, i.e.
one attributes the phase of the point on the limit cycle, 
to which asymptotically the given point in
the basin of attraction converges -- thus the term ``asymptotic''.

The definition of the phase can be straightforwardly generalised to the case of 
many stable limit  
cycles $\mathbf{x}^{lc(1)},
\mathbf{x}^{lc (2)},\ldots$. Each of the phases $\varphi^{(k)}$ is defined in the 
corresponding basin of attraction
$U^{(k)}$. All the cycles have generically   different frequencies $\omega^{(k)}$, however, 
the values of the frequencies 
are not relevant for the definition of the phases.
The phases are not defined on the basin boundaries, and on the 
sets which do not belong to the basins.

\section{Phase Response Curve and Phase Transfer Curve}
\label{sec:prctrc}
 
Phase Response Curve (PRC) describes effect of a pulse force (kick) 
on the phase of the oscillations. One supposes that
an action of a pulse can be described as a map 
$\mathbf{x}\to \bar{\mathbf{x}}=Q(\mathbf{x})$, and that the both points
lie in the same basin. If the initial point
lies on the limit cycle, it is characterized by the phase $\varphi$, 
and we get a mapping to the new phase  
\[
\varphi\to \bar{\varphi}=\text{PRC}(\varphi)=\Phi(Q(\mathbf{x}^{lc}(\varphi)))\;.
\]

For a single kick action, the PRC provides a full information about the phase shift 
as a result of the kick. If several kicks 
are applied (or a regular or an irregular sequence of kicks), the PRC is useful, 
if the interval between the kicks is 
large enough. Indeed, the PRC is defined for the points on the limit cycle, 
and to apply it to subsequent kicks, one
needs to ensure that prior to a kick the point is on the limit cycle, or at 
least very close to it. This means, that the
time interval between the kicks should be larger than the relaxation time 
from the point $Q(\mathbf{x}^{lc})$ to the
limit cycle. If the time interval between the kicks is 
short (or the relaxation time is large), 
one should take into account
corrections as described in 
Refs.~\cite{Zeng-Glass-Shrier-92,Krishnan-Bazhenov-Pikovsky-13,Klinshov_etal-17}. 

For a given sequence of kicks $Q_n$ (for generality, one can assume that all 
the kicks are different)
occurring at time instants $t_n$, one can write a phase evolution maps, provided
that the time intervals $t_{n+1}-t_n$ between the kicks are large enough:
\begin{equation}
\varphi_{n+1}=\omega (t_{n+1}-t_n)+\text{PRC}_n(\varphi_n)\;.
\label{eq:pdprc}
\end{equation}
The dynamics of this one-dimensional map, which can be not one-to-one, describes 
different effects of synchronization and chaotization of
periodic oscillations by an external pulse force.

For several coexisting limit cycle oscillations, the phase response curves can be 
defined for each of them:
\[
\varphi^{(k)}\to \bar{\varphi}^{(k)}=\text{PRC}^{(k)}(\varphi^{(k)})\;.
\]
 However, now there is a possibility that the state after the kick $\bar{\mathbf{x}}$ 
 belongs to the 
 basin of another cycle, i.e. the kick switches from a source periodic regime to 
a target one. Still, we can define the relation
 between the new and the old phases via the \textit{Phase Transfer Curve} PTC:
\[
\varphi^{(k)}\to \bar{\varphi}^{(j)}=\text{PTC}^{(k\to j)}(\varphi^{(k)})\;.
\]
Schematically we illustrate kicks leading to PRC and PTC in Fig.~\ref{fig:isc}
below.

%\begin{figure}[!thb]
%\includegraphics[width=0.4\columnwidth]{fig1.png}
%\caption{Schematic illustration of two kicks leading to PRC (the point after the kick
%is in the basin of cycle 1) and to a PTC (the point after the kick is
%in the basin of cycle 2). The basin boundary is a dashed line. } 
%\label{fig:prcptc}
%\end{figure}

Neither PRC nor PTC are defined for the states that are mapped by the kick to the 
basin boundaries. However, if these 
boundaries are simple (fixed points, unstable limit cycles), 
PRC and PTC are not defined at a 
finite set of the phases. Close to this set, the relaxation time from the kicked state  
$\bar{\mathbf{x}}=Q(\mathbf{x})$ to the corresponding
stable limit cycle is large (it diverges as the
image point $\bar{\mathbf{x}}$ approaches the basin boundary). Therefore, 
there are small regions where the 
phase approximation leading to the dynamics of type~\eqref{eq:pdprc} is 
not valid. The size of these regions decreases
with the increase of the time intervals between the kicks $t_{n+1}-t_n$.

 In the next section we will apply the approach based on PRC and PTC 
 to a particular system with two stable limit cycles.
 
 \section{Bistable Stuart-Landau-type oscillator}
 
 \subsection{Phase dynamics}
 Here we illustrate the approach with a generalization of the Stuart-Landau oscillator:
 \begin{equation}
\begin{aligned}
\dot{x} & =  -y \cdot \Phi(x, y) -x \cdot R(x, y) \;,\\
\dot{y} & = x \cdot\Phi(x, y) - y \cdot R(x, y)\;,
\end{aligned}
\label{eq:bsl1}
\end{equation}
where
\[
R(x, y) =\beta (x^2 + y^2 -a^2)\cdot(x^2 + y^2 -b^2) \cdot (x^2 + y^2-c^2),\quad
\Phi (x, y) = \omega+ \gamma(x^2 + y^2).
\]
These equations have a simple form if written in polar coordinates
$
x=r\cos\theta,\quad y=r\sin\theta
$:
\begin{equation}
\begin{aligned}
\dot r&=-\beta r(r^2-a^2)(r^2-b^2)(r^2-c^2)\;,\\
\dot\theta&=\omega+\gamma r^2\;.
\end{aligned}
\label{eq1}
\end{equation}
Below we assume $\beta>0$ and $a<b<c$. Then one can easily see that the system possesses
two stable limit cycles: cycle 1 with $r=a$ and $\dot\theta=\Omega_1=\omega+\gamma a^2$, 
and
cycle 2 with $r=c$ and $\dot\theta=\Omega_2=\omega+\gamma c^2$. The basin boundary
separating basins 
of these two cycles is the unstable cycle $r=b$. 

We now introduce the two phases in the corresponding
basins. In the basin of cycle 1,
i.e. for $0<r<b$, the phase $\varphi^{(1)}$ should fulfil $\dot \varphi^{(1)}=\Omega_1$.
We look for a representation 
\begin{equation}
\varphi^{(1)}=\theta+f_1(r)\;.
\label{eq:ph1}
\end{equation}
Taking the time derivative of $\varphi^{(1)}$, we obtain for the function $f_1$
the following equation
\[
\Omega_1=\omega+\gamma a^2=
\omega+\gamma r^2+\frac{df_1}{dr}[-\beta r(r^2-a^2)(r^2-b^2)(r^2-c^2)]\;.
\]
Integration (with a condition that on the cycle 1 $\varphi^{(1)}$ coincides with $\theta$)
yields
\begin{equation}
f_1(r)=\frac{\gamma\ln\frac{|c^2-r^2|}{|c^2-a^2|}}{2\beta c^2(c^2-b^2)}-
\frac{\gamma\ln\frac{|b^2-r^2|}{|b^2-a^2|}}{2\beta b^2(c^2-b^2)}+
\frac{\gamma\ln\frac{r}{a}}{\beta b^2c^2}\;.
\label{eq:f1}
\end{equation}
Similarly, in the basin of cycle 2 $r>b$, we define the phase
$\varphi^{(2)}$ satisfying $\dot \varphi^{(2)}=\Omega_2$, which can be represented as
\begin{equation}\varphi^{(2)}=\theta+f_2(r)
\label{eq:ph2}
\end{equation}
with
\begin{equation}
f_2(r)=\frac{\gamma\ln\frac{|a^2-r^2|}{|a^2-c^2|}}{2\beta a^2(a^2-b^2)}-
\frac{\gamma\ln\frac{|b^2-r^2|}{|b^2-c^2|}}{2\beta b^2(a^2-b^2)}+
\frac{\gamma\ln\frac{r}{c}}{\beta b^2a^2}\;.
\label{eq:f2}
\end{equation}

\begin{figure}[!thb]
\includegraphics[width=0.5\columnwidth]{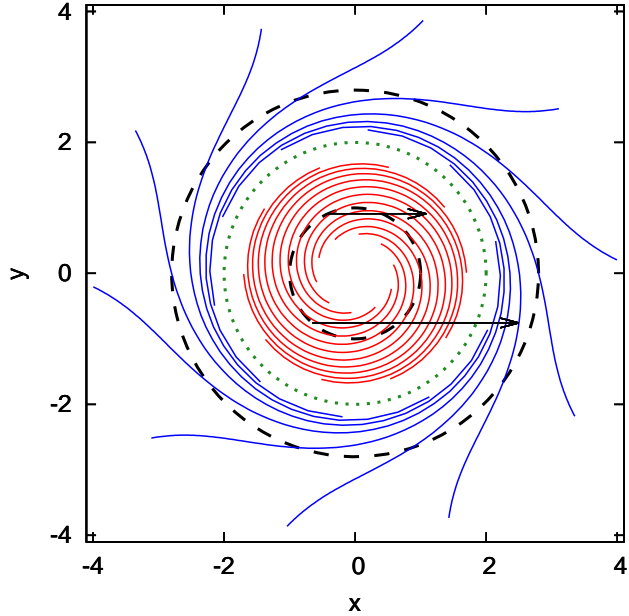}
\caption{Phase space of the bistable Stuart-Landau oscillator
for $a=1,b=2,c=2.8,\gamma=1,\beta=0.01$. 
Dashed lines: stable cycles; green dots: unstable cycle. 
Red lines: isochrons of cycle 1; blue lines: isochrons of cycle 2.
Black arrows show two kicks: one moves a point on the cycle 1 to the basin
of the same cycle (this kick leads to a PRC); 
another one moves a point on the cycle 1
to the basin of the cycle 2 (this kick leads to a PTC).} 
\label{fig:isc}
\end{figure}

The isochrons are the lines of constant phases. In polar coordinates they are represented by
families of curves $\theta=\varphi-f_{1,2}(r)$. We illustrate the isochrons for the bistable
Stuart-Landau oscillator in Fig~\ref{fig:isc}.

\subsection{PRC and PTC for the bistable Stuart-Landau oscillator}
Now we derive the PRC and the PTC for the oscillator~\eqref{eq:bsl1}. We assume that 
the external action is a kick with strength $A$ in $x$-direction: $x\to x+A$. Consider a point
on the cycle 1 with the phase $\varphi^{(1)}$. In the 
polar coordinates the point just after the kick is
\[
\tilde r=\sqrt{(a\cos\varphi^{(1)}+A)^2+a^2\sin^2\varphi^{(1)}},\qquad 
\tilde\theta=\text{ATAN2}(a\sin\varphi^{(1)},a\cos\varphi^{(1)}+A)
\;.\]
This point lies in the basin of cycle 1 if $\tilde r<b$, otherwise it belongs to the basin of cycle 2.
Thus, according to expressions \eqref{eq:ph1},\eqref{eq:ph2}, the new phases are
\begin{align}
\varphi^{(1)}_{new}&=\tilde\theta+f_1(\tilde r)\qquad\text{if}\quad \tilde r<b \;,\label{eq:prc11}\\
\varphi^{(2)}_{new}&=\tilde\theta+f_2(\tilde r)\qquad\text{if}\quad \tilde r>b\;.\label{eq:prc12}
\end{align}
Similar expressions describe the target phase if the source point is on cycle 2:
\begin{align}
\varphi^{(1)}_{new}&=\bar\theta+f_1(\bar r)\qquad\text{if}\quad \bar r<b\;,\label{eq:prc21}\\
\varphi^{(2)}_{new}&=\bar\theta+f_2(\bar r)\qquad\text{if}\quad \bar r>b\;,\label{eq:prc22}
\end{align}
where
\[
\bar r=\sqrt{(c\cos\varphi^{(2)}+A)^2+c^2\sin^2\varphi^{(2)}},\qquad 
\bar\theta=\text{ATAN2}(c\sin\varphi^{(2)},c\cos\varphi^{(2)}+A)
\;.\]
These expressions give the analytic forms of $\text{PRC}^{(1\to1)}$ \eqref{eq:prc11},
of $\text{PTC}^{(1\to2)}$ \eqref{eq:prc12}, of $\text{PRC}^{(2\to2)}$ \eqref{eq:prc22},
and of $\text{PTC}^{(2\to1)}$ \eqref{eq:prc21}.

\begin{figure}[!thb]
\includegraphics[width=0.6\columnwidth]{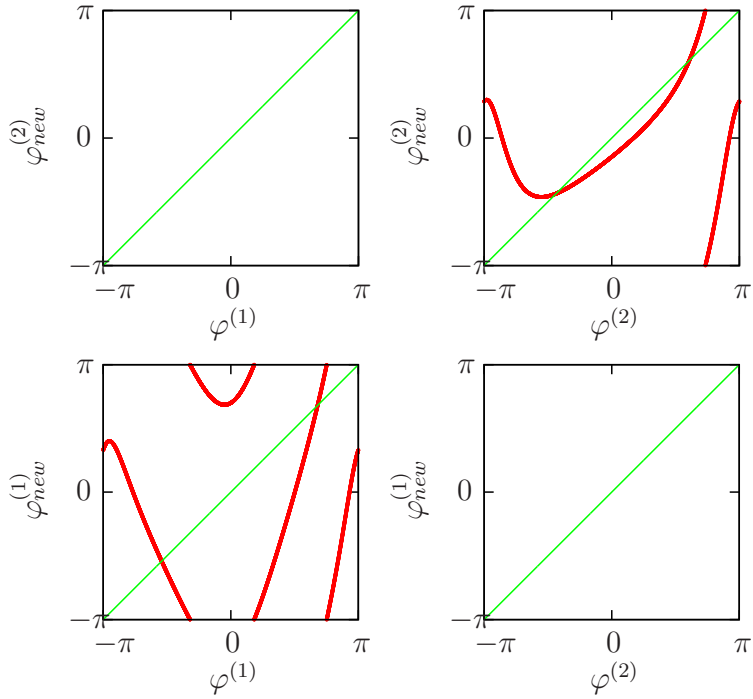}
\caption{PRC and PTC for the same parameters as in Fig.~\ref{fig:isc} 
and $A=0.67$. \tcb{Two diagonal 
panels show the PRCs 
$\vp^{(1)}\to\vp^{(1)}_{new}$ and $\vp^{(2)}\to\vp^{(2)}_{new}$.
Two non-diagonal panels that should show PTCs ($\vp^{(1)}\to\vp^{(2)}_{new}$ 
and $\vp^{(2)}\to\vp^{(1)}_{new}$) are in fact empty for this small
forcing (no transition from one cycle to another one); 
however the same  plots for larger forcing in 
Figs.~\ref{fig:prc90} and \ref{fig:prc150} do show nontrivial PTCs.}} 
\label{fig:prc67}
\end{figure}

\begin{figure}[!thb]
\includegraphics[width=0.6\columnwidth]{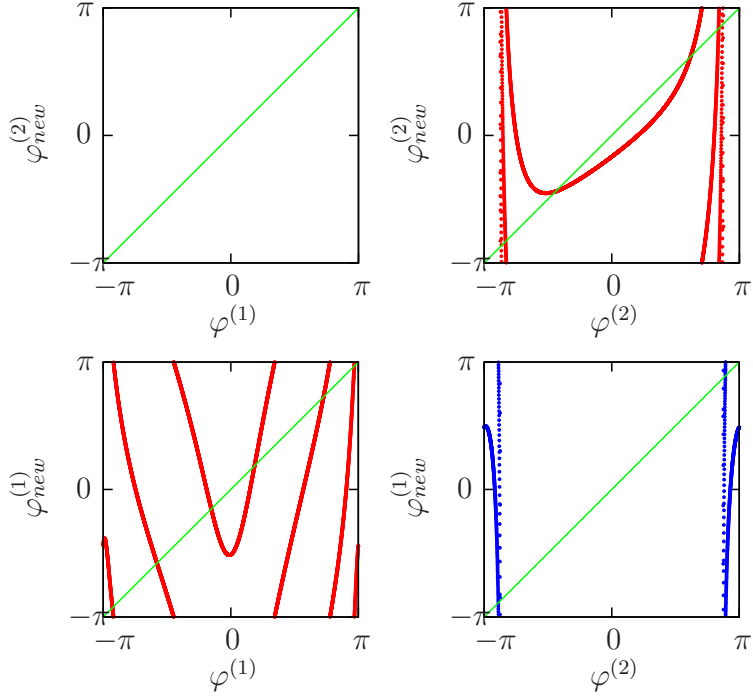}
\caption{The same as in Fig.~\ref{fig:prc67}, but for $A=0.9$.
For this moderate amplitude of the kicks also the PTC
$\vp^{(2)}\to\vp^{(1)}_{new}$ exists.} 
\label{fig:prc90}
\end{figure}

\begin{figure}[!thb]
\includegraphics[width=0.6\columnwidth]{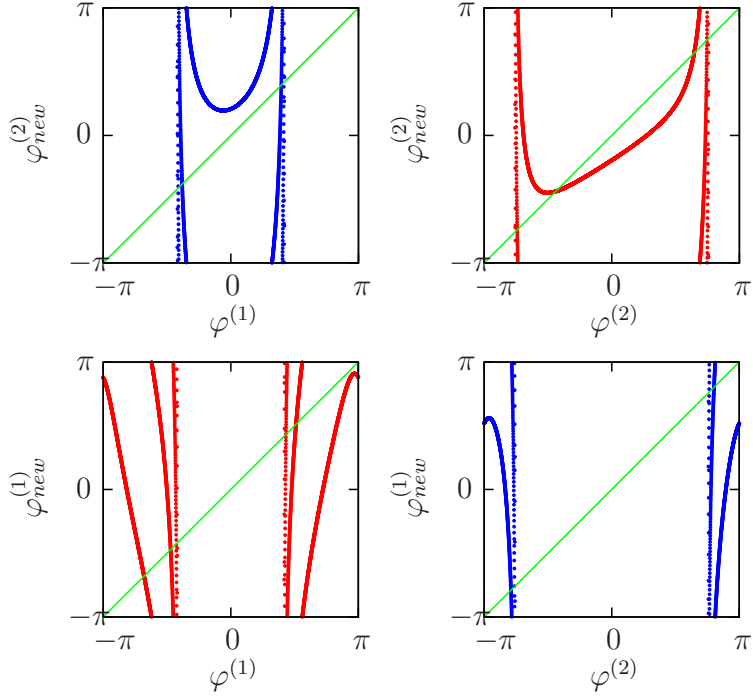}
\caption{The same as in Fig.~\ref{fig:prc67}, but for $A=1.5$.
For this large amplitude of the kicks all PRCs and PTCs exist.} 
\label{fig:prc150}
\end{figure}

We illustrate  these curves in Figs.~\ref{fig:prc67},\ref{fig:prc90},\ref{fig:prc150}. 
Fig.~\ref{fig:prc67} shows a case of small kick amplitudes, where there is no transition
between the cycles. As a result, only PRCs exist, and there are no PTCs. For an intermediate 
kick amplitude, there is a possibility for a transition from cycle 2 
to cycle 1, but not in the opposite direction
(Fig.~\ref{fig:prc90}). Finally, for a large kick amplitude, all 
transitions are possible and all PRCs and PTCs
exist (Fig.~\ref{fig:prc150}). 

An important feature of PTCs and PRCs in the case where PTCs exist, is the 
singularity of these curves at the boundaries of domains of their definition. The analytic
nature of these singularities is determined by terms $\sim \log|r^2-b^2|$ in expressions
\eqref{eq:f1},\eqref{eq:f2}. The points which are mapped to a vicinity of the basin boundary
$r=b$, spend a logarithmically long time before they are attracted to stable cycles 1 or 2,
and during this time interval, an extreme sensitivity of the final phase with respect to the initial 
one is reached. Formally, there is a phase at which both the PRC and the PTC are not defined, this 
is the point which is mapped exactly on the basin boundary.

\section{Periodically kicked bistable oscillator}

\subsection{Validity of one-dimensional approximation}
In the previous section we derived the PRCs and PTCs that determine the
shift of the phases under a single kick. Here we will apply 
them to a periodically kicked oscillator
\begin{equation}
\begin{aligned}
\dot{x} & =  -y \cdot \Phi(x, y) -x \cdot R(x, y) +A\sum_n\delta(t-nT)\;,\\
\dot{y} & = x \cdot\Phi(x, y) - y \cdot R(x, y)\;.
\end{aligned}
\label{eq:bsl2}
\end{equation}
In the case of very large period $T$, one can assume that just prior to the next kick, the 
state of the system is very close to one of the stable cycles. Thus,  the
dynamics can be described with the one-dimensional 
mappings 
\begin{align*}
\varphi^{(1)}_{n+1}=\text{PRC}^{(1\to1)}(\varphi^{(1)}_n)& \text{ if }r_n=a,\tilde r_n<b\;,\\
\varphi^{(2)}_{n+1}=\text{PTC}^{(1\to 2)}(\varphi^{(1)}_n)& \text{ if }r_n=a,\tilde r_n>b\;,\\
\varphi^{(2)}_{n+1}=\text{PRC}^{(2\to2)}(\varphi^{(2)}_n)&\text{ if }r_n=c,\bar r_n>b\;,\\
\varphi^{(1)}_{n+1}=\text{PTC}^{(2\to 1)}(\varphi^{(2)}_n)&\text{ if }r_n=c,\bar r_n<b\;.\\
\end{align*}
This description will be not so perfect for small periods $T$, because in this case deviations
from the limit cycles prior to the next kick will be significant.

To illustrate this, we first derive the exact two-dimensional mapping, with which 
the one-dimensional
mapping based on PRCs and PTCs will be compared.
Let us denote $r_n,\varphi_n$ the state just prior to the $n$-th kick. Then
$\theta_n=\varphi_n-f_{1,2}(r_n)$ and $x_n=r_n\cos\theta_n$, $y_n=r_n\sin\theta_n$.
Just after 
the kick we have
\[
\hat x=x_n+A,\quad \hat y=y_n,\quad \hat r=\sqrt{(x_n+A)^2+y_n^2},\quad 
\hat \theta=\text{ATAN2}(y_n,x_n+A)\;.
\]
The evolution of this point during time $T$ can be calculated 
using the equation for $r$:
\[
\dot r=-\beta r(r^2-a^2)(r^2-b^2)(r^2-c^2)\;.
\]
Integration of this equation with starting point $\hat r$ yields
\begin{gather*}
\frac{\ln \left|\frac{r_{n+1}^2-c^2}{\hat r^2-c^2}\right|}{c^6+a^2b^2c^2-c^4(a^2+b^2)}+
\frac{\ln \left|\frac{r_{n+1}^2-b^2}{\hat r^2-b^2}\right|}{b^6+a^2b^2c^2-b^4(a^2+c^2)}+\\+
\frac{\ln \left|\frac{r_{n+1}^2-a^2}{\hat r^2-a^2}\right|}{a^6+a^2b^2c^2-a^4(b^2+c^2)}-
\frac{\ln\left|\frac{r_{n+1}^2}{\hat r^2}\right|}{a^2b^2c^2}=-2\beta T
\end{gather*}
Unfortunately, from this equation it is hardly possible to find the function 
$r_{n+1}(T,\hat r)$ explicitly.
This function can, however, be found numerically
 as a root of a function of one variable.

Depending on in which basin the point $\hat r$ lies, we then 
find $\varphi_{n+1}$ from the
following expressions:
\[
\varphi_{n+1}=\begin{cases} \hat\theta+T\Omega_2+f_2(\hat r)& \hat r>b\;, \\
\hat\theta+T\Omega_1+f_1(\hat r)&\hat r<b\;.
\end{cases}
\]

\begin{figure}[!thb]
\includegraphics[width=0.8\columnwidth]{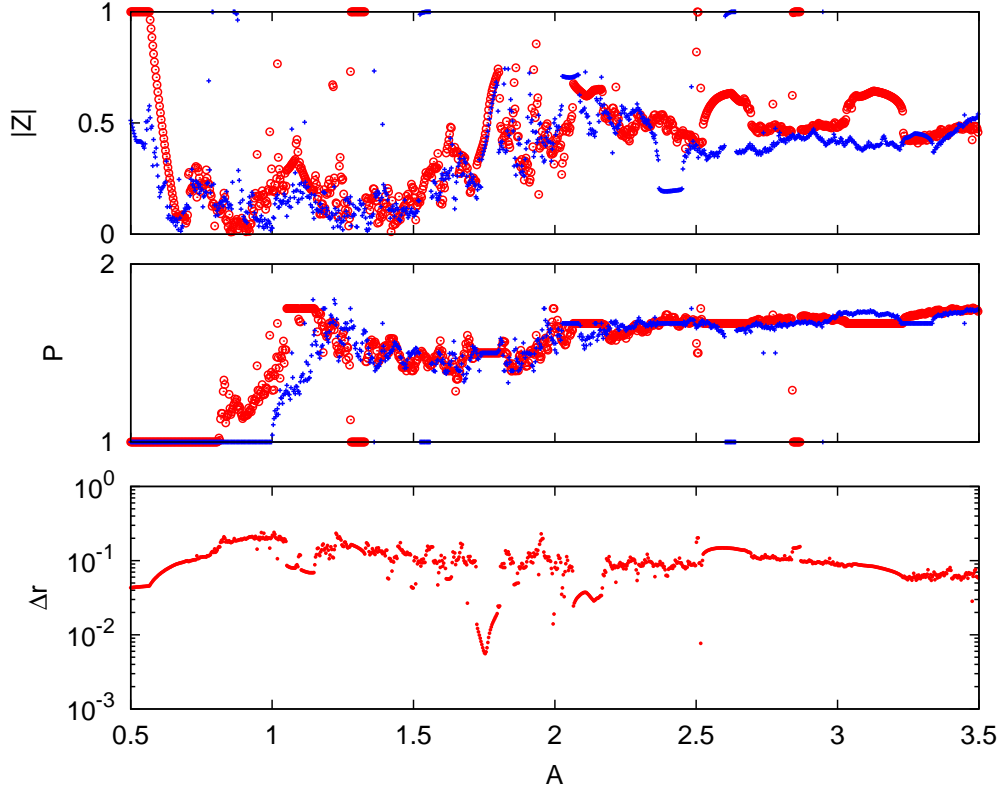}
\caption{Comparison of the statistical properties of the dynamics (as described in the text)
in the full two-dimensional system (red circles)
with the one-dimensional map (blue pluses), for $T=4$. In these simulations 
the starting point was close
to cycle 1, therefore bistability of attractors at small values of $A$ is not
revealed.} 
\label{fig:c04}
\end{figure}

\begin{figure}[!thb]
\includegraphics[width=0.8\columnwidth]{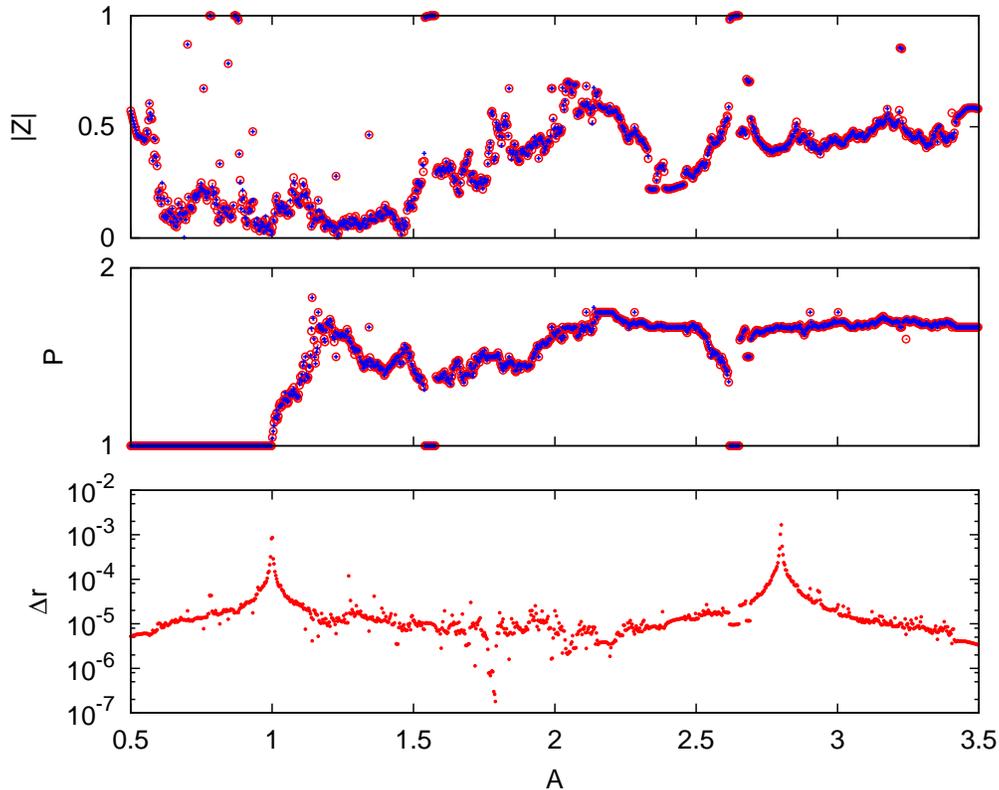}
\caption{The same as Fig.~\ref{fig:c04}, but for $T=27$.} 
\label{fig:c27}
\end{figure}

We now compare the results of simulations of the exact two-dimensional mapping derived, 
with the simulations based on the one-dimensional approximation via PRCs and PTCs.
In Fig.~\ref{fig:c04} we show the case of small period $T=4$, in Fig.~\ref{fig:c27} 
the period is large $T=27$.
In both cases we show, as functions of the kick  parameter $A$, three 
averaged quantities. Quantity
$|Z|$ characterizes the distribution of the phases, here the 
complex ``order parameter'' of
the distribution of the phases is defined as $Z=\langle \exp[i\varphi_n]\rangle$. 
If the state of the system is a stable 
fixed point, then $|Z|=1$, otherwise $|Z|<1$ (this quantity, however, does not allow 
distinguishing between 
regular and chaotic states). Quantity $P$ describes distribution of the 
points between the two basins 
of attraction, it is calculated as $\langle \text{ind}_n\rangle$, where 
ind=1 in the basin of cycle 1 and
ind=2 in the basin of cycle 2. This quantity allows distinguishing 
regimes belonging to one basin only, and
those with switchings between the cycles 1 and 2. Finally, for the two-dimensional map
we characterize deviations from the stable cycles via quantity 
$\Delta r=\langle (r_n-q_n)^2\rangle^{1/2}$,
where $q_n=a$ if $\text{ind}_n=1$, and $q_n=c$ if $\text{ind}_n=2$.

Let us discuss first the quantity $\Delta r$, as it mostly directly 
characterizes quality of the 
one-dimensional
approximation. One can see that for $T=27$ (Fig.~\ref{fig:c27}) the typical 
values are $10^{-5}$, what means
that
here the one-dimensional map is rather close to the exact one. One can also see that
the approximation is bad for special values of kick amplitude $A\approx a=1$ 
and $A\approx c=2.8$. The reason is that at these special values of $A$, 
points from the stable cycles
are mapped exactly on the origin, from the vicinity of which a trajectory 
only slowly evolves toward the
attracting cycle 1. For $T=4$ (Fig.~\ref{fig:c04}) the characteristic values 
of $\Delta r$ are much larger, around $10^{-1}$,
here one cannot expect the one-dimensional approximation to work well. 
This is indeed evident 
from the inspection of the average characteristics $|Z|$ and $P$. 
For $T=4$ their values
in the exact solution and in the one-dimensional approximation 
differ significantly, while
for $T=27$ they practically coincide. 
\tcb{In Figs.~\ref{fig:c04},\ref{fig:c27} we illustrated dependence on the
kick amplitude $A$ for the two selected values of time interval $T$. A more thorough
study of $T$-dependence shows that the values of $\Delta r$ drastically depend
on whether the regime is chaotic (like for cases $T=4$ and $T=27$ presented in
 Figs.~\ref{fig:c04},\ref{fig:c27}), or periodic. In the latter case the values
 of $\Delta r$ can be as small as the accuracy of numerics $\sim 10^{-12}$. If 
 one excludes such periodic cases, the decrease in the value of $\Delta r$
 (averaged over the values of kick amplitudes in the same range as in 
 Figs.~\ref{fig:c04},\ref{fig:c27}) follows an exponential
 law $\langle \Delta r\rangle_A\sim 10^{-0.18 T -0.2}$.}

\subsection{Structure of chaos}
\begin{figure}[!thb]
\includegraphics[width=0.6\columnwidth]{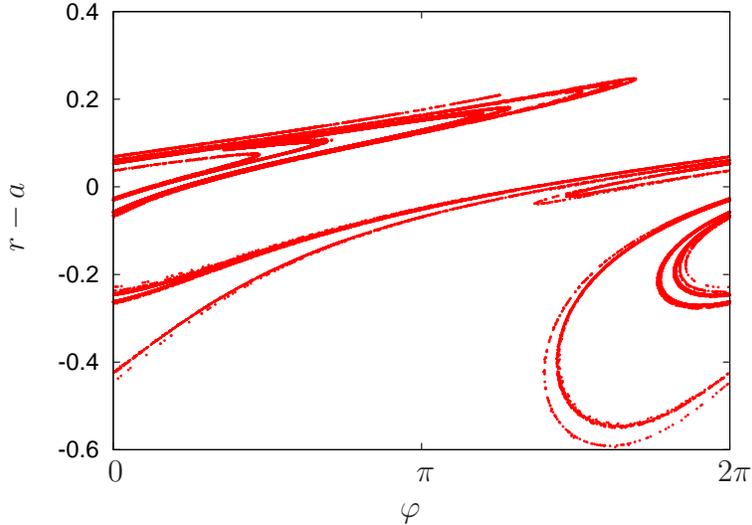}
\caption{The strange attractor close to cycle 1 for $T=4$ and $A=0.8$.} 
\label{fig:attr2}
\end{figure}

At many parameters of forcing the bistable oscillator~\eqref{eq:bsl2} 
demonstrates chaos. If the kick amplitude is small, there are no transitions from one
basin to another, and the properties of chaos are similar to that of the kicked
standard monostable Stuart-Landau oscillator~\cite{Zaslavsky-78}. We illustrate this
regime in Fig.~\ref{fig:attr2}, showing the attractor near the cycle 1 for $T=4$ and $A=0.8$.
One can clearly see fractal set of stripes, typical for chaotic two-dimensional sets.

\begin{figure}[!thb]
\includegraphics[width=\columnwidth]{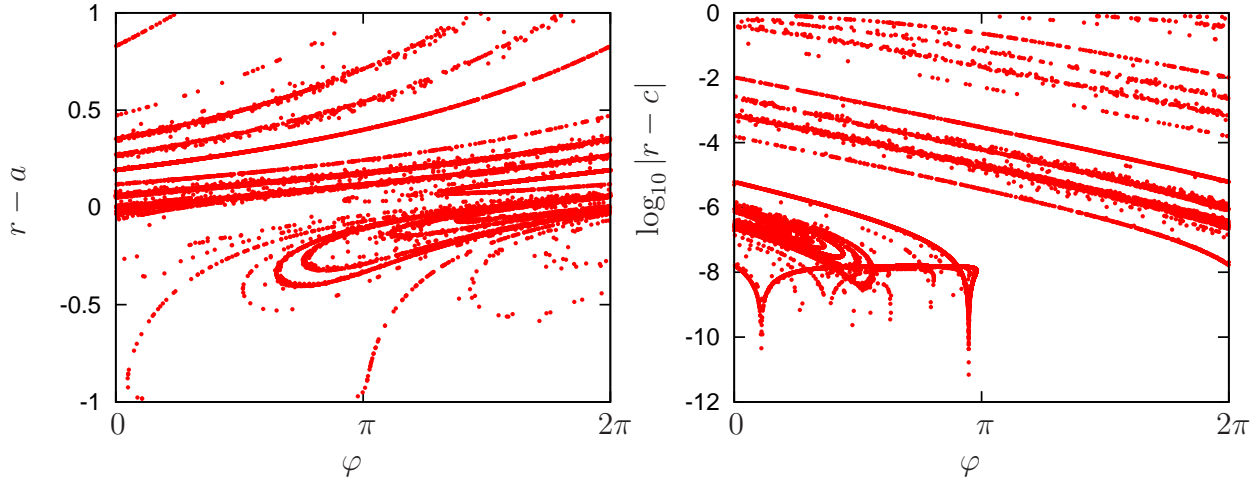}
\caption{The strange attractor in the basin of cycle 1 (left panel) and in 
the basin of cycle 2 (right panel)
 for $T=4$ and $A=1.4$. Stripes on the right panel cross because the 
 absolute value is used for the observable shown.} 
\label{fig:attr1}
\end{figure}

The structure of chaos in the case where there are transitions between the basins 
(parameters $T=4$, $A=1.4$),
Fig.~\ref{fig:attr1}, is more complex. Here we show a neighborhood of cycle 1 
in usual coordinates,
while to reveal a fine structure close to cycle 2 
we plot $\log_{10}|r-c|$ vs $\varphi$. One can see
that the fractal structure is somehow smeared: 
together with well-defined stripes there are also
``scattered'' points. The reason for this are extreme variations of local 
expansion/contraction 
rates at the transitions between the basins. In the one-dimensional approximation
these variations are manifested by singularities of the PRCs and PTCs at the boundaries 
of their definition. Points of the attractor that are mapped due to a kick to a vicinity of
the unstable cycle at $r=b$, are extremely scattered; to reveal a fine 
fractal structure of these scattered 
sets one needs very long trajectories.

There is also another peculiarity of the system \eqref{eq:bsl2}: the two stable cycles 
for the parameters chosen have very 
different stability properties: the  multiplier of cycle 1 is $0.42$, while 
for cycle 2 it is $0.072$.
Therefore, close to cycle 2 the convergence is very strong, and already for moderate 
time intervals between the kicks $T$, distance of points to cycle 2 reaches machine zero 
($\approx 10^{-14}$) for double 
precision calculations. One can see in   Fig.~\ref{fig:attr1}, that 
already for $T=4$ a typical
distance of the attractor from the cycle 2 is $\sim 10^{-8}$. For $T\approx 8$, the 
machine accuracy is reached,
and no reliable calculations with double precision of the structure 
of an attractor close to cycle 2 are possible.

\section{Conclusion}
Summarizing, in this paper we have generalized the concept 
of phase response curves on the case of multistable periodic oscillators.
The transitions from one basin to another one can be described by the
phase transfer curves. We presented a solvable example, where both phase
response curves and phase transfer curves can be found analytically.
Furthermore, we checked how good is the phase approximation, based solely
on the PRCs and PTCs, in comparison with exact solutions where the variations 
of the amplitudes are not neglected. \tcb{ While for the illustration only
the solvable model of a bistable Start-Landau-type oscillator have been
used, we do not see any restriction in application of the concept to other
systems with different types of the phase dynamics and of the phase sensitivity 
(e.g., to relaxation oscillations). Furthermore, the 
methods of experimental determination of a PRC~\cite{Galan-Ermentrout-Urban-05}
can be straightforwardly extended to PTC determination as well.}

\acknowledgments
We acknowledge support  by the Russian 
Science Foundation (Grant 17-12-01534). 

\def\cprime{$'$}

\end{document}